# Cardiac SPECT Radiomics Features Repeatability and Reproducibility: A Multi Scanner Phantom Study

Mohammad Edalat-Javid[1], Isaac Shiri[2], Ghasem Hajianfar[3], Hamid Abdollahi[4], Niki Oveisi[5], Mohammad Javadian[6], Mojtaba Shamsaei Zafarghandi[1], Hadi Malek[7], Ahmad Bitarafan-Rajabi [7,8,9], Mehrdad Oveisi[2,10]

1. Faculty of Nuclear Engineering and Physics, Amirkabir University of Technology, Tehran, Iran
2. Division of Nuclear Medicine and Molecular Imaging, Department of Medical Imaging, Geneva University Hospital, CH-1211 Geneva 4, Switzerland
3. Rajaie Cardiovascular Medical and Research Center, Iran University of Medical Science, Tehran, Iran
4. Department of Radiologic Sciences and Medical Physics, Faculty of Allied Medicine, Kerman University, Kerman, Iran
5. School of Population and Public Health, The University of British Columbia, BC, V6T 1Z4, Canada.
6. Department of Computer Science, Kermanshah University of Technology, Kermanshah, Iran
7. Department of Nuclear Medicine and Molecular Imaging, Rajaie Cardiovascular, Medical, and Research Center, Iran University of Medical Sciences, Tehran, Iran
8. Cardiovascular Intervention Research Center, Rajaie Cardiovascular Medical and Research Center, Iran University of Medical Sciences, Tehran, Iran
9. Echocardiography Research Center, Rajaie Cardiovascular Medical and Research Center, Iran University of Medical Sciences, Tehran, Iran
10. Department of Computer Science, University of British Columbia, Vancouver BC, Canada

## Corresponding Author: Isaac Shiri

Division of Nuclear Medicine and Molecular Imaging,
Department of Medical Imaging, Geneva University Hospital,
CH-1211 Geneva 4, Switzerland
Email: Isaac.sh92@gmail.com

# Abstract

**Background**: The aim of this study was to assess the robustness of cardiac SPECT radiomics features against changes in imaging settings including acquisition and reconstruction settings.

**Methods**: Four scanners were used to acquire SPECT scans of a cardiac phantom with 5mCi of $^{99m}$Tc. The effects of different image acquisition and reconstruction settings including the Number of View, View Matrix Size, attenuation correction, image reconstruction algorithm, number of iterations, number of subsets, type of filter, full width at half maximum (FWHM) of Gaussian filter, Butterworth filter order, and Butterworth filter cut-off were studied. In total 5263 different images were reconstructed. Eighty-seven radiomic features including first, second, and high order textures were extracted from images. To assess reproducibility and repeatability the coefficient of variation (COV) was used for each image feature over the different imaging settings.

**Result:** IDMN and IDN features from GLCM, RP from GLRLM, ZE from GLSZM, and DE from GLDM feature sets were the only features that were the most reproducible (COV ≤ 5%) against changes in all imaging settings. In addition, the IDMN feature from GLCM, LALGLE, SALGLE and LGLZE from GLSZM, and SDLGLE from GLDM feature sets were the features that were less reproducible (COV>20 %) against changes in all imaging settings. Matrix size has the greatest impact on feature variability as most of features are not repeatable and 82.76% of them had (COV>20 %).

**Conclusion:** Repeatability and reproducibility of SPECT/CT radiomics texture features in different imaging settings is feature-dependent, and different image acquisitions and reconstructions have different effects on radiomics texture features. Low COV radiomics features could be consider for further clinical studies.

## Introduction

As major causes of death, cardiovascular diseases (CVDs) are main concerns for many scientists worldwide (1, 2). Myocardial perfusion imaging (MPI) is a valuable approach to identify CVDs patients for medical management such as diagnosis, intervention, therapy, and follow-up (3). With this regards, nuclear medicine modalities including single photon emission computed tomography (SPECT) remains the most common procedure in the evaluation and risk stratification of patients with known or suspected CVDs (4). Studies have indicated that SPECT and SPECT-CT have high diagnostic accuracy, low radiation exposure, and high image quality for CVD management (5). Advances in cardiac nuclear medicine imaging in terms of software and hardware such as optimal detector geometric arrays, linear count statistics, count rate response, and new reconstruction algorithms, provide further improvement in image quality (6, 7).

Recently, quantitative radiomics studies have opened new horizons for better medical management of several diseases such as cancer and CVDs (8-13). The aim of radiomics is to extract quantitative features from medical images using data-mining algorithms for predicting, prognosis, and therapeutic response prediction and assessment (8, 14, 15). In this light, radiomics could provide valuable information for personalized therapy. Previous radiomics studies have suggested that radiomics features could act as biomarkers that characterize and predict diseases to provide support for patient management (8, 16).

Based on biomarker discovery guidelines and studies, biomarker repeatability and reproducibility are critical and essential assessments that should be addressed prior to clinical decision making. (17). In the repeatability and reproducibility measurements, a reliable radiomic feature remains stable between two measurements when conditions remain stable. The feature should also remain the same while using different equipment, software, settings, or operator (18). If so, then the feature may be considered as a good biomarker for clinical settings. Due to this, a considerable amount of literature has been published on radiomics features repeatability and reproducibility against changes in the radiomics process such as image acquisition, reconstruction, pre-processing, segmentation, and data analysis (19-21). Nuclear radiomics studies have tested the repeatability and reproducibility of imaging features over various imaging parameters including reconstruction algorithms, matrix size, iteration number, number of subsets, and post-filtering in both phantom and patients(18, 21).

To date, little evidence has been found on cardiac SPECT repeatability and reproducibility over different imaging settings. This present study aims to assess the repeatability and reproducibility of radiomics features for cardiac phantoms against variations in image acquisitions and reconstruction methods.

## Material and Methods:

### Strategy of Study

Fig 1 shows the details of the current study.

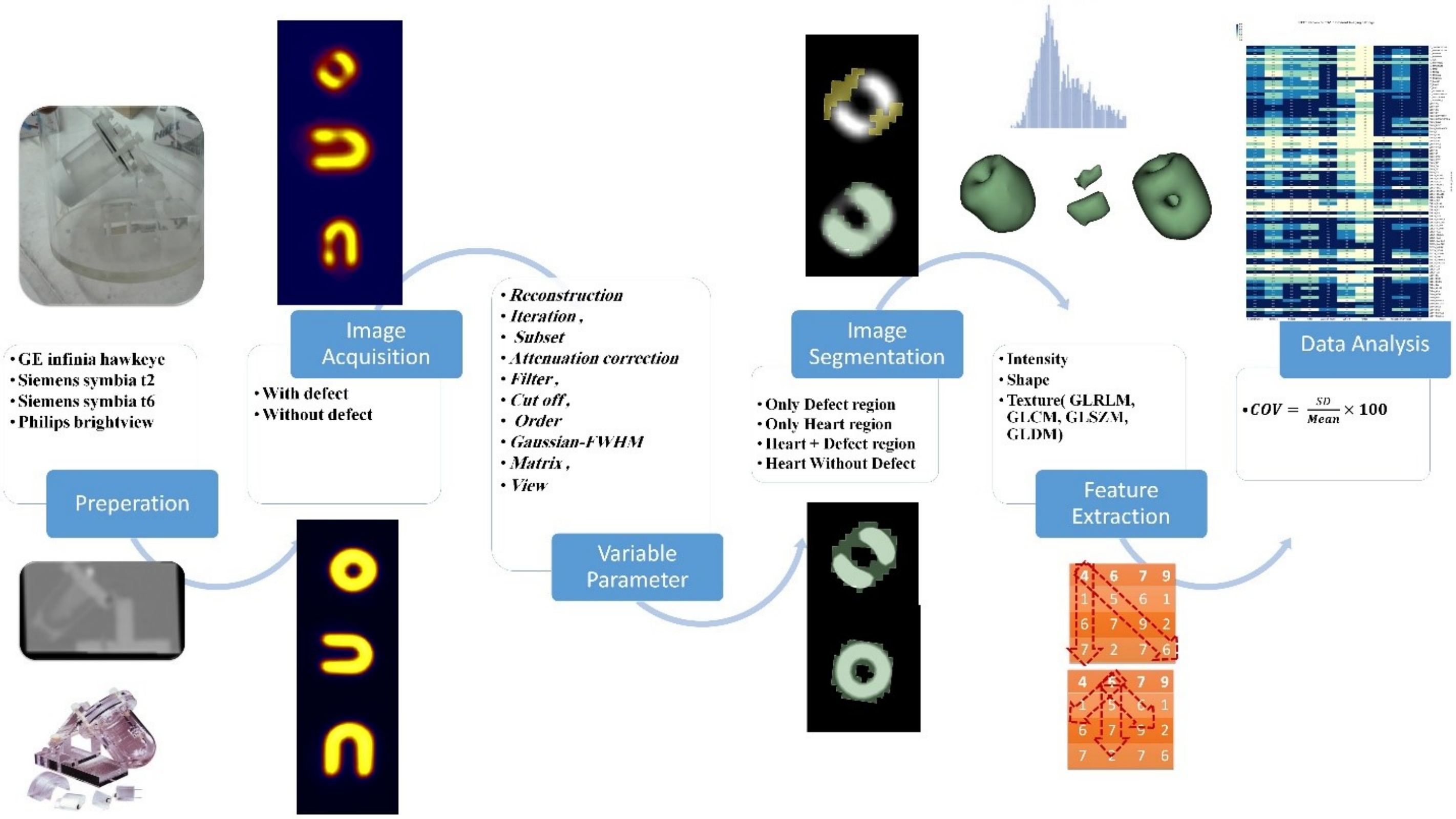

**Figure 1.** Illustrates the process flow followed in the paper.

### Scanners

Four scanners with dual-head gamma cameras were used to acquire SPECT scans of a phantom with cardiac defects. Clinical data was obtained using three SPECT/CT scanners (GE INFINIA HAWKEYE, SIEMENS SYMBIA T2, SIEMENS SYMBIA T6, and PHILIPS BRIGHTVIEW)

### Phantom Preparation

A commercially available phantom mimicking the shape of a normal heart was used in these experiments. The right ventricular cavity was filled with a solution of water, -Sestamibi (MIBI), and only 5mCi of $^{99m}$Tc to avoid any saturation-related loss in counts. This phantom was placed in the Jaszczak Phantom and was surrounded by water. In order to simulate the real position of the cardiac in the chest, the Jaszczak Phantom was placed in the center of the field of view and orientated in the 45 left-anterior and 45 caudal directions.

**Data acquisition**

Time per projection was modulated to obtain a total recorded activity of approximately 500 kilo-counts. Three defects were also added in order to simulate clinical abnormalities. Data acquisitions were performed at different scanners with different acquisition modes such as Number of View, View Matrix Size, and attenuation correction.

**Image Reconstruction**

To study the impact of reconstruction settings on image features, six image reconstruction methods were used: Filter Back Projection (FBP), Ordered Subset Expectation Maximization (OSEM), Maximum Likelihood Expectation Maximization (MLEM) reconstruction, FLASH 3D, ASTONISH, and WALLIS. The effects of different settings including number of iterations, number of subsets, different filter (Butterwort, Hanning, Metz, Shepp Logan, Gaussian, Parzan), full width at half maximum (FWHM) of Gaussian filter, Butterworth filter order, and Butterworth filter cut-off were also studied. All these parameters are listed in Table 1 and resulted in 5263 reconstructed images.

**Image Segmentation**

All segmentations were performed using the 3D-Slicer software. For the non-defected cardiac, the whole cardiac was segmented. For the defected cardiac, three regions were extracted including the defect region, whole cardiac, and whole cardiac minus defect area. To minimize the impact of segmentation on the results, one VOI was delineated and registered on all reconstruction methods.

**Table 1**. The variable and constant parameters

| Parameter studied | variable | stable |
|---|---|---|
| **Reconstruction algorithm** | FBP , OSEM , FLASH 3D, ASTONISH, MLEM, WALLIS | Iteration=2 , Subset=8, Filter=BW ,Cutoff=.5 , Order=10 , Matrix=64 ,View=64 |
| **Iteration** | 1, 2, 3, 4, 6, 8, 10, 12, 14, 16, 25 | FWHM= 5mm , View=32 , Matrix=64 |
| **Subset** | 2, 4, 8, 16 | Iteration=2 , FWHM=5mm<br>Matrix=64 , view=32 |
| **Filter (FWHM in mm)** | 0, 0.5, 1, 1.5, 2, 2.5, 3, 3.5, 4, 4.5, 5, 5.5, 6, 6.5, 7 | Iteration=2 , Subset=8<br>Matrix=64 , View=64 |
| **FILTER** | Butterwort, hanning, metz, shepp logan, gussian, sheep logan, par | Matrix=64 , View=64<br>cutoff=.5 , Order=10 |
| **CUTOFF** | .35, .4, .45, .5, .55 | Matrix=64 , View=32<br>Filter=BW , Order=10 |
| **ORDER** | 1.5, 1.75, 2, 5, 9, 10, 20, 30 | Matrix=64 , View=32<br>Filter=BW , Cut off=.5 |
| **Attenuation Correction** | Device type | Matrix=64 , View=64 ,Filter=BW , cutoff=.5 , Order=10 |
| **MATRIX** | 64, 128, 256 | View=64 , Filter=BW<br>Cutoff=.5 , Order=5 |
| **VIEW** | 32, 64, 128 | Matrix=64 , Filter=BW<br>Cut off=.5 , Order=5 |

**Feature Extraction**

Eighty-seven radiomic features including first, second, and high order texture were extracted from images. Table 2 shows the extracted image features.

**Statistical Analysis**

To assess reproducibility and repeatability the coefficient of variation (COV) was used for each image feature over different imaging settings, by:

$$COV = \frac{SD}{Mean} \times 100$$

Where SD is the standard deviation of feature values and Mean is the mean of different settings. COVs were analyzed and four reproducibility categories were obtained based on the COV values: very small (COV ≤ 5%), small (5% < COV ≤ 10%), intermediate (10% < COV ≤ 20%) and large (COV > 20%).

**Table 2.** The radiomics features

| **First Order Statistics (FOS)** | **Gray Level Co-occurrence Matrix (GLCM)** | **Gray Level Run Length Matrix (GLRLM)** |
|---|---|---|
| 1. Energy | 1. Autocorrelation(AC) | 1. Short Run Emphasis (SRE) |
| 2. Total Energy | 2. Joint Average(JA) | 2. Long Run Emphasis (LRE) |
| 3. Entropy | 3. Cluster Prominence(CP) | 3. Gray Level Non-Uniformity (GLN) |
| 4. Minimum | 4. Cluster Shade(CS) | 4. Gray Level Non-Uniformity Normalized (GLNN) |
| 5. 10th percentile | 5. Cluster Tendency(CT) | 5. Run Length Non-Uniformity (RLN) |
| 6. 90th percentile | 6. Contrast | 6. Run Length Non-Uniformity Normalized (RLNN) |
| 7. Maximum | 7. Correlation | 7. Run Percentage (RP) |
| 8. Mean | 8. Difference Average(DAve) | 8. Gray Level Variance (GLV) |
| 9. Median | 9. Difference Entropy(DEnt) | 9. Run Variance (RV) |
| 10. Interquartile Range( IQR) | 10. Difference Variance(DVariance) | 10. Run Entropy (RE) |
| 11. Range | 11. Joint Energy(JEne) | 11. Low Gray Level Run Emphasis (LGLRE) |
| 12. Mean Absolute Deviation (MAD) | 12. Joint Entropy(JEnt) | 12. High Gray Level Run Emphasis (HGLRE) |
| 13. Robust Mean Absolute Deviation (RMAD) | 13. Informal Measure of Correlation (IMC) 1 | 13. Short Run Low Gray Level Emphasis (SRLGLE) |
| 14. Root Mean Squared (RMS) | 14. Informal Measure of Correlation (IMC) 2 | 14. Short Run High Gray Level Emphasis (SRHGLE) |
| 15. Skewness | 15. Inverse Difference Moment (IDM) | 15. Long Run Low Gray Level Emphasis (LRLGLE) |
| 16. Kurtosis | 16. Inverse Difference Moment Normalized (IDMN) | 16. Long Run High Gray Level Emphasis (LRHGLE) |
| 17. Variance | 17. Inverse Difference (ID) | **Gray Level Dependence Matrix (GLDM)** |
| 18. Uniformity | 18. Inverse Difference Normalized (IDN) | 1. Small Dependence Emphasis (SDE) |
| | 19. Inverse Variance(IV) | 2. Large Dependence Emphasis (LDE) |
| | 20. Maximum Probability(MP) | 3. Gray Level Non-Uniformity (GLN) |
| | 21. Sum Average(SA) | 4. Dependence Non-Uniformity (DN) |
| | 22. Sum Entropy(SE) | 5. Dependence Non-Uniformity Normalized (DNN) |
| | 23. Sum of Squares(SS) | 6. Gray Level Variance (GLV) |
| | **Gray Level Size Zone Matrix (GLSZM)** | 7. Dependence Variance (DV) |
| | 1. Small Area Emphasis (SAE) | 8. Dependence Entropy (DE) |
| | 2. Large Area Emphasis (LAE) | 9. Low Gray Level Emphasis (LGLE) |
| | 3. Gray Level Non-Uniformity (GLN) | 10. High Gray Level Emphasis (HGLE) |
| | 4. Gray Level Non-Uniformity Normalized (GLNN) | 11. Small Dependence Low Gray Level Emphasis (SDLGLE) |
| | 5. Size-Zone Non-Uniformity (SZN) | 12. Small Dependence High Gray Level Emphasis (SDHGLE) |
| | 6. Size-Zone Non-Uniformity Normalized (SZNN) | 13. Large Dependence Low Gray Level Emphasis (LDLGLE) |
| | 7. Zone Percentage (ZP) | 14. Large Dependence High Gray Level Emphasis (LDHGLE) |
| | 8. Gray Level Variance (GLV) | |
| | 9. Zone Variance (ZV) | |
| | 10. Zone Entropy (ZE) | |
| | 11. Low Gray Level Zone Emphasis (LGLZE) | |
| | 12. High Gray Level Zone Emphasis (HGLZE) | |
| | 13. Small Area Low Gray Level Emphasis (SALGLE) | |
| | 14. Small Area High Gray Level Emphasis (SAHGLE) | |
| | 15. Large Area Low Gray Level Emphasis (LALGLE) | |
| | 16. Large Area High Gray Level Emphasis (LAHGLE) | |

**Results:**

Figure 2 depicts the heatmap of radiomics features in different imaging settings based on COV values: 1: very small (COV ≤ 5%), 2: small (5% < COV ≤ 10%), 3: intermediate (10% < COV ≤ 20%) and 4: large (COV > 20%). Table 3 provides the percentage of different COV groups in the different imaging settings.

**Impact of reconstruction, number of iterations and number of subsets**

For reconstruction, 16.90% (14 features) and 42.53% (37 features) of all features were found as most (COV ≤ 5%), and less reproducible (COV > 20%), respectively. Details on these features are available in supplementary Table 1. Most of the less reproducible features were from GLRLM, GLSZM, and GLDM feature sets. The most reproducible features against reconstruction are 2 features of FO, 7 features of GLCM, 5 features of GLRLM, 2 features of GLSZM, and a feature of GLDM; as seen in supplementary Table 1. These features are 10Percentile/Entropy (from the FO feature sets), CS/ IDMN/IDN/Imc1/Imc2/JENT/SE (from the GLCM feature sets), LRE/RE/RLNUN/RP/SRE (from the GLRLM feature sets), SZE/ZP (from the GLSZM feature sets), and DE (from the GLDM feature sets).

On the impact of the number of iterations, it was found that 28.74% of all features had COV ≤ 5% (25 features). Features including Entropy/Kurtosis/Mean/Median/RMS (from the FO feature sets), CS/DENT/IDMN/IDN/IMC1/IMC2/JENE/JENT/SE (from the GLCM feature sets), LRE/RE/RLNU/RLNUN/RP/SRE (from the GLRLM feature sets), SZNUN/SAE/ZE/ZP (from the GLSZM feature sets) and DE/SDE (from the GLDM feature sets) had the highest reproducibility (COV ≤ 5%). From the GLCM and FO feature sets, just one feature was found as less reproducible (COV>20%). These features were Minimum and CP from the FO and GLCM feature sets, respectively. More details are available in supplementary Table 2.

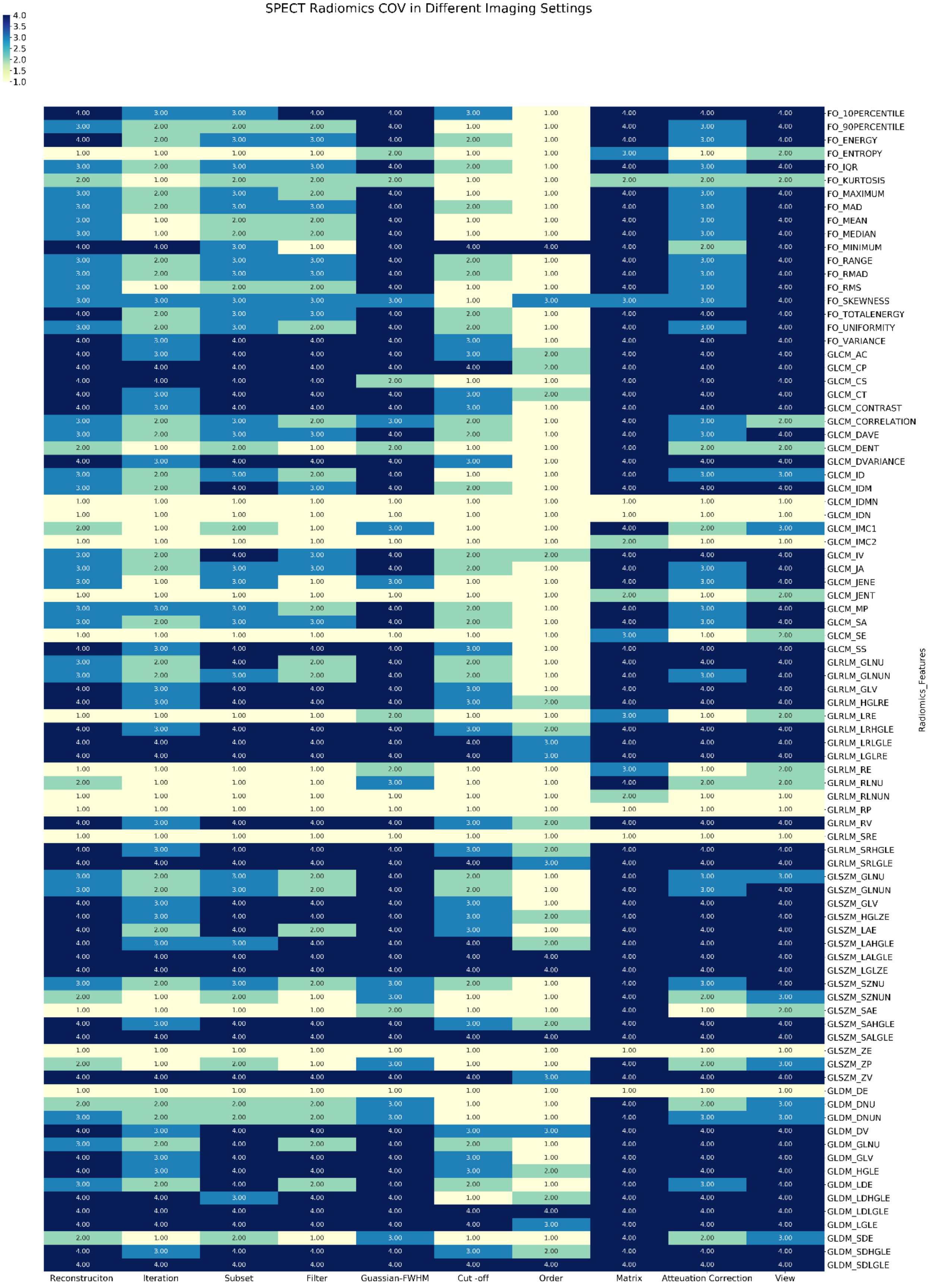

**Figure 2.** SPECT radiomics COV in different Imaging setting

In regards to the impact of the number of subsets on feature reproducibility, 17.24% (15 features) and 41.38% (36 features) of all features had COV ≤ 5% and COV>20% respectively. From FO feature set, Variance was the less reproducible feature (COV>20%) and from GLDM, only DE was the most reproducible feature (COV>20%). In addition, features including Entropy Minimum, CS, IDMN, IDN, IMC1, IMC2, JENT, SE, LRE, RE, RLNU, RLNUN, RP, SRE, SAE, and ZE from different feature sets had a COV ≤ 5% and were introduced as the most reproducible feature. On the other hand, features AC, CP, CT, Contrast, DVARIANCE, IDM,

IV, SS, GLNU, GLV, HGLRE, LRHGLE, LRLGLE, LGLRE, RV, SRHGLE, SRLGLE, GLV, HGLZE, LAE, LALGLE, LGLZE, SAHGLE, SALGLE, ZV, DV, GLNU, GLV, HGLE, LDE, LDLGLE, LGLE, SDHGLE, and SDLGLE had the highest variations against change in the number of subsets (COV>20%). More details are available in supplementary Table 3.

**Impacts of Different Filter, FWHM of Gaussian filter, and Cut–off and Order of Butterworth filter**

Results on the impact of filter showed that 22.59% of all features (22 FEATURES), had COV≤5% and features including 10percentile/Entropy/Minimum (from the FO feature set), DENT/IDMN/IDN/IMC1/IMC2/JENE/JENT/SE (from the GLCM feature set), LRE/RE/RLNU/RLNUN/RP/SRE (from the GLRLM feature set), SZNUN/SAE/ZE/ZP (from the GLSZM feature set), and DE/SDE (from the GLDM feature set) were the most reproducible features. Of the less reproducible features, 33.79% of all features (33 features) had COV>20% and several features from GLCM, GLRLM, GLSZM, and GLDM feature sets were found as less reproducible (COV>20%). More details are available in supplementary Table 4.

On the impact of Gaussian-FWHM, results showed that 11.49% (10 features) and 67.82% (59 features) of all features had COV≤5% and COV>20%, respectively. Interestingly, from the FO,

GLSZM, and GLDM features sets, just one feature was found as most reproducible (COV ≤ 5%). These features were 10 percentile (FO), ZE (GLSZM) and DE (GLDM). On the other hand, a wide range of features was found as less reproducible (COV>20%). These features are detailed in supplementary Table 5.

Results for order showed that 37.93% (33 features) and 68.97% (60 features) of all features were most reproducible (COV ≤ 5%), respectively. Interestingly for GLCM and GLRLM feature sets, there was no less reproducible feature (COV>20%). For FO features, just Skewness and Minimum were not reproducible (COV>5%). In addition, for GLSZM and GLDM feature sets, features including LALGLE/LGLZE/SALGLE and LDLGLE/SDLGLE had COV>20%. More details are available in supplementary Table 6.

Regarding the cut off, results showed that 37.93% (33 features) of all features were most reproducible (COV≤5%). For GLCM and FO feature sets, CP and Minimum were the less reproducible (COV>20 %) feature, respectively. More details are available in supplementary Table 7.

**Impact of Matrix size**

On the impact of matrix size, data shows that most features are not repeatable and 82.76% of them had COV>20 %. On the other hand, only seven features including IDMN

IDN/IMC1/ (from GLCM), RP/SRE, (from GLRLE), ZE (from GLSZM), and DE (GLDM) were found as most reproducible (COV≤5%) and there were no reproducible features from the FO feature set. More details are available in supplementary Table 8.

**Impact of Attenuation Correction**

The results on the impact of attenuation correction showed that 16.09% of all features (16 features) had COV≤5%. These features are 10percentile/Entropy (from the FO feature set),

CS/IDMN/IDN/IMC1/IMC2/JENT/SE (from the GLCM feature set), LRE/RE/RLNUN/RP/SRE (from GLRLM feature set), SAE/ZE (from the GLSZM feature set), and DE (from the GLDM feature set). In addition, 44.83% (39 features) of all features were less reproducible. More details are available in supplementary Table 9.

**Impact of number of views**

On the impact of the number of views, results showed that 70.11% (61 features) and 9.2% (8 features) of all features had COV≤5% and COV>20%, respectively. Features including IDMN/IDN/IMC1/IMC2 (from the GLCM feature set), RLNUN/RP/SRE (from the GLRLM feature set), ZE (from the GLSZM feature set), and DE (from the GLDM feature set) were the most reproducible features (COV≤5%) and there was feature with COV≤5% in FO feature set. More details are available in supplementary Table 10.

**Table 3.** The Percent of different COV group in different imaging setting

| COV | Reconstruction | Iteration | Subset | Filter | Gaussian-FWHM | Cut -off | Order | Matrix | Attenuation Correction | Number of View |
|---|---|---|---|---|---|---|---|---|---|---|
| COV ≤ 5% | 16.09 | 28.74 | 17.24 | 25.29 | 11.49 | 37.93 | 68.97 | 6.9 | 16.09 | 9.2 |
| 5% < COV ≤ 10% | 9.2 | 29.89 | 13.79 | 22.99 | 8.05 | 24.14 | 16.09 | 4.6 | 10.34 | 11.49 |
| 10% < COV ≤ 20% | 32.18 | 25.29 | 27.59 | 13.79 | 12.64 | 22.99 | 8.05 | 5.75 | 28.74 | 9.2 |
| COV > 20% | 42.53 | 16.09 | 41.38 | 37.93 | 67.82 | 14.94 | 6.9 | 82.76 | 44.83 | 70.11 |

**Supplemental Table 1.** Effect of reconstruction

| **Feature category** | **Feature parent** | **COV≤5%** | **5%<COV≤10%** | **10%<COV≤20%** | **COV>20%** |
|---|---|---|---|---|---|
| First order | FO | 10Percentile/ Entropy | Kurtosis | 90Percentile/ IQR/ Maximum/ MAD/ Mean/ Median/ Range/ RMAD/ RMS/ Skewness/ Uniformity | Energy/ Minimum/ TotalEnergy/ Variance |
| Second order | GLCM | CS/ Idmn/ Idn/ Imc1/ Imc2/ Jent/ SE | DEnt | Correlation/ Dave/ Id/ Idm/ IV/ JA/ JEne/ MP/ SA | AC/ CP/ CT/ Contrast/ DVariance/ SS |
| Higher order | GLRLM | LRE/ RE/ RLNUN/ RP/ SRE | RLNU | GLNU/ GLNUN | GLV/ HGLRE/ LRHGLE/ LRLGLE/ LGLRE/ RV/ SRHGLE/ SRLGLE |
| | GLSZM | SZE/ ZP | SZNUN/ ZP | GLNU/ GLNUN/ SZNU | GLV/ HGLZE/ LAE/ LAHGLE/ LALGLE/ LGLZE/ SAHGLE/ SALGLE/ ZV |
| | GLDM | DE | DNU/ SDE | DNUN/ GLNU/ LDE | DV/ GLV/ HGLE/ LDHGLE/ LDLGLE/ LGLE/ SDHGLE/ SDLGLE |

**Supplemental Table 2.** Effect of iteration

| Feature category | Feature parent | COV≤5% | 5%<COV≤10% | 10%<COV≤20% | COV>20% |
|---|---|---|---|---|---|
| First order | FO | Entropy/ Kurtosis/ Mean/ Median/ RMS | 90Percentile/ Energy/ IQR/ Maximum/ MAD/ Range/ RMAD/ TotalEnergy/ Uniformity | 10Percentile<br>Skewness<br>Variance | Minimum |
| Second order | GLCM | CS/ DEnt/ Idmn/ Idn/ Imc1/ Imc2/ JEne/ Jent/ SE | Correlation<br>DAve<br>Id<br>Idm/ IV/ JA/ SA | AC<br>CT<br>Contrast<br>DVariance<br>MP<br>SS | CP |
| Higher order | GLRLM | LRE/ RE/ RLNU/ RLNUN/ RP/ SRE | GLNU<br>GLNUN | GLV<br>HGLRE<br>LRHGLE<br>RV<br>SRHGLE | LRLGLE<br>LGLRE<br>SRLGLE |
| | GLSZM | SZNUN/ SAE/ ZE/ ZP | GLNU<br>GLNUN<br>LAE<br>SZNU | GLV<br>HGLZE<br>LAHGLE<br>SAHGLE | LALGLE<br>LGLZE<br>SALGLE<br>ZV |
| | GLDM | DE/ SDE | DNU<br>DNUN<br>GLNU<br>LDE | DV<br>GLV<br>HGLE<br>SDHGLE | LDHGLE<br>LDLGLE<br>LGLE<br>SDLGLE |

**Supplemental Table 3.** Effect of subset

| Feature category | Feature parent | COV≤5% | 5%<COV≤10% | 10%<COV≤20% | COV>20% |
|---|---|---|---|---|---|
| First order | FO | Entropy<br>Minimum | 90Percentile<br>Kurtosis<br>Mean<br>Median<br>RMS | 10Percentile<br>Energy<br>IQR<br>Maximum<br>MAD<br>Range<br>RMAD<br>Skewness<br>TotalEnergy<br>Uniformity | Variance |
| Second order | GLCM | CS<br>Idmn<br>Idn<br>Imc1<br>Imc2<br>Jent<br>SE | DEnt | Correlation<br>DAve<br>Id<br>JA<br>JEne<br>MP<br>SA | AC<br>CP<br>CT<br>Contrast<br>DVariance<br>Idm<br>IV<br>SS |
| Higher order | GLRLM | LRE<br>RE<br>RLNU<br>RLNUN<br>RP<br>SRE | | GLNUN | GLNU<br>GLV<br>HGLRE<br>LRHGLE<br>LRLGLE<br>LGLRE<br>RV<br>SRHGLE<br>SRLGLE |
| | GLSZM | SAE<br>ZE | SZNUN<br>ZP | GLNU<br>GLNUN<br>LAHGLE<br>SZNU | GLV<br>HGLZE<br>LAE<br>LALGLE<br>LGLZE<br>SAHGLE<br>SALGLE<br>ZV |
| | GLDM | DE | DNU<br>DNUN<br>SDE | LDHGLE | DV<br>GLNU<br>GLV<br>HGLE<br>LDE<br>LDLGLE<br>LGLE<br>SDHGLE<br>SDLGLE |

**Supplemental Table 4.** Effect of filter

| Feature category | Feature parent | COV≤5% | 5%<COV≤10% | 10%<COV≤20% | COV>20% |
|---|---|---|---|---|---|
| First order | FO | 10Percentile<br>Entropy<br>Minimum | 90Percentile<br>Kurtosis<br>Maximum<br>Mean<br>Median<br>RMS<br>Uniformity | Energy<br>IQR<br>MAD<br>Range<br>RMAD<br>Skewness<br>TotalEnergy | Variance |
| Second order | GLCM | DEnt<br>Idmn<br>Idn<br>Imc1<br>Imc2<br>JEne<br>Jent<br>SE | Correlation<br>Id<br>MP | DAve<br>Idm<br>IV<br>JA<br>SA | AC<br>CP<br>CS<br>CT<br>Contrast<br>DVariance<br>SS |
| Higher order | GLRLM | LRE<br>RE<br>RLNU<br>RLNUN<br>RP<br>SRE | GLNU<br>GLNUN | | GLV<br>HGLRE<br>LRHGLE<br>LRLGLE<br>LGLRE<br>RV<br>SRHGLE<br>SRLGLE |
| | GLSZM | SZNUN<br>SAE<br>ZE<br>ZP | GLNU<br>GLNUN<br>LAE<br>SZNU | | GLV<br>HGLZE<br>LAHGLE<br>LALGLE<br>LGLZE<br>SAHGLE<br>SALGLE<br>ZV |
| | GLDM | DE<br>SDE | DNU<br>DNUN<br>GLNU<br>LDE | | DV<br>GLV<br>HGLE<br>LDHGLE<br>LDLGLE<br>LGLE<br>SDHGLE<br>SDLGLE |

**Supplemental Table 5.** Effect of Guassian-FWHM

| Feature category | Feature parent | COV≤5% | 5%<COV≤10% | 10%<COV≤20% | COV>20% |
|---|---|---|---|---|---|
| First order | FO | 10Percentile | Entropy<br>Kurtosis | Skewness | 90Percentile<br>Energy<br>IQR<br>Maximum<br>MAD<br>Mean<br>Median<br>Minimum<br>Range<br>RMAD<br>RMS<br>TotalEnergy<br>Uniformity<br>Variance |
| Second order | GLCM | CS<br>Idmn<br>Idn<br>Imc1<br>Imc2<br>Jent<br>SE | DEnt | Correlation<br>JEne | AC<br>CP<br>CT<br>Contrast<br>DAve<br>DVariance<br>Id<br>Idm<br>IV<br>JA<br>MP<br>SA<br>SS |
| Higher order | GLRLM | RLNUN<br>RP<br>SRE | LRE<br>RE | RLNU | GLNU<br>GLNUN<br>GLV<br>HGLRE<br>LRHGLE<br>LRLGLE<br>LGLRE<br>RV<br>SRHGLE<br>SRLGLE |
| | GLSZM | ZE | SAE | SZNU<br>SZNUN<br>ZP | GLNU<br>GLNUN<br>GLV<br>HGLZE<br>LAE<br>LAHGLE<br>LALGLE<br>LGLZE<br>SAHGLE<br>SALGLE<br>ZV |
| | GLDM | DE | DNU<br>DNUN<br>GLNU<br>LDE | DNU<br>DNUN<br>SDE | DV<br>GLNU<br>GLV<br>HGLE<br>LDE<br>LDHGLE<br>LDLGLE<br>LGLE<br>SDHGLE<br>SDLGLE |

**Supplemental Table 6.** Effect of cut off

| Feature category | Feature parent | COV≤5% | 5%<COV≤10% | 10%<COV≤20% | COV>20% |
|---|---|---|---|---|---|
| First order | FO | 90Percentile<br>Entropy<br>Kurtosis<br>Maximum<br>Mean<br>Median<br>RMS<br>Skewness | Energy<br>IQR<br>MAD<br>Range<br>RMAD<br>TotalEnergy<br>Uniformity | 10Percentile<br>Variance | Minimum |
| Second order | GLCM | CS<br>DEnt<br>Id<br>Idmn<br>Idn<br>Imc1<br>Imc2<br>JEne<br>Jent<br>SE | Correlation<br>DAve<br>Idm<br>IV<br>JA<br>MP<br>SA | AC<br>CT<br>Contrast<br>DVariance<br>SS | CP |
| Higher order | GLRLM | LRE<br>RE<br>RLNU<br>RLNUN<br>RP<br>SRE | GLNU<br>GLNUN | GLV<br>HGLRE<br>LRHGLE<br>RV<br>SRHGLE | LRLGLE<br>LGLRE<br>SRLGLE |
| | GLSZM | SZNUN<br>SAE<br>ZE<br>ZP | GLNU<br>GLNUN<br>SZNU | GLV<br>HGLZE<br>LAE<br>SAHGLE | LAHGLE<br>LALGLE<br>LGLZE<br>SALGLE<br>ZV |
| | GLDM | DE<br>DNU<br>DNUN<br>LDHGLE<br>SDE | GLNU<br>LDE | DV<br>GLV<br>HGLE<br>SDHGLE | LDLGLE<br>LGLE<br>SDLGLE |

**Supplemental Table 7.** Effect of order

| Feature category | Feature parent | COV≤5% | 5%<COV≤10% | 10%<COV≤20% | COV>20% |
|---|---|---|---|---|---|
| First order | FO | 10Percentile<br>90Percentile<br>Energy<br>Entropy<br>IQR<br>Kurtosis<br>Maximum<br>MAD<br>Mean<br>Median<br>Range<br>RMAD<br>RMS<br>TotalEnergy<br>Uniformity<br>Variance | | Skewness | Minimum |
| Second order | GLCM | CS<br>Contrast<br>Correlation<br>DAve<br>DEnt<br>DVariance<br>Id<br>Idm<br>Idmn<br>Idn<br>Imc1<br>Imc2<br>JA<br>JEne<br>Jent<br>MP<br>SA<br>SE<br>SS | AC<br>CP<br>CT<br>IV | | |
| Higher order | GLRLM | GLNU<br>GLNUN<br>GLV<br>LRE<br>RE<br>RLNU<br>RLNUN<br>RP<br>SRE | HGLRE<br>LRHGLE<br>RV<br>SRHGLE | LRLGLE<br>LGLRE<br>SRLGLE | |
| | GLSZM | GLNU<br>GLNUN<br>GLV<br>LAE<br>SZNU<br>SZNUN<br>SAE<br>ZE<br>ZP | HGLZE<br>LAHGLE<br>SAHGLE | ZV | LALGLE<br>LGLZE<br>SALGLE |
| | GLDM | DE<br>DNU<br>DNUN<br>GLNU<br>GLV<br>LDE<br>SDE | HGLE<br>LDHGLE<br>SDHGLE | DV<br>LGLE | LDLGLE<br>SDLGLE |

## **Supplemental Table 8.** Effect of matrix

| Feature category | Feature parent | COV≤5% | 5%<COV≤10% | 10%<COV≤20% | COV>20% |
|---|---|---|---|---|---|
| First order | FO | | Kurtosis | Entropy<br>Skewness | 10Percentile<br>90Percentile<br>Energy/ IQR<br>Maximum<br>MAD/ Mean<br>Median<br>Minimum<br>Range/ RMAD<br>RMS<br>TotalEnergy<br>Uniformity<br>Variance |
| Second order | GLCM | Idmn<br>Idn<br>Imc1 | Imc2<br>Jent | SE | AC/ CP/ CS/ CT<br>Contrast<br>Correlation<br>Dave/ DEnt<br>DVariance<br>Id/ Idm/ IV/ JA<br>JEne/ MP/ SA/ SS |
| Higher order | GLRLM | RP<br>SRE | RLNUN | LRE<br>RE | GLNU<br>GLNUN<br>GLV<br>HGLRE<br>LRHGLE<br>LRLGLE<br>LGLRE<br>RLNU<br>RV<br>SRHGLE<br>SRLGLE |
| | GLSZM | ZE | | | GLNU<br>GLNUN<br>GLV<br>HGLZE<br>LAE<br>LAHGLE<br>LALGLE<br>LGLZE<br>SZNU<br>SZNUN<br>SAE<br>SAHGLE<br>SALGLE<br>ZP<br>ZV |
| | GLDM | DE | | | DNU<br>DNUN<br>DV<br>GLNU<br>GLV<br>HGLE<br>LDE<br>LDHGLE<br>LDLGLE<br>LGLE<br>SDE<br>SDHGLE<br>SDLGLE |

**Supplemental Table 9.** Effect of view

| Feature category | Feature parent | COV≤5% | 5%<COV≤10% | 10%<COV≤20% | COV>20% |
|---|---|---|---|---|---|
| First order | FO | | Entropy<br>Kurtosis | | 10Percentile<br>90Percentile<br>Energy<br>IQR<br>Maximum<br>MAD<br>Mean<br>Median<br>Minimum<br>Range<br>RMAD<br>RMS<br>Skewness<br>TotalEnergy<br>Uniformity<br>Variance |
| Second order | GLCM | Idmn<br>Idn<br>Imc1<br>Imc2 | Correlation<br>DEnt<br>Jent<br>SE | Id | AC<br>CP<br>CS<br>CT<br>Contrast<br>DAve<br>DVariance<br>Idm<br>IV<br>JA<br>JEne<br>MP<br>SA<br>SS |
| Higher order | GLRLM | RLNUN<br>RP<br>SRE | LRE<br>RE<br>RLNU | | GLNU<br>GLNUN<br>GLV<br>HGLRE<br>LRHGLE<br>LRLGLE<br>LGLRE<br>RV<br>SRHGLE<br>SRLGLE |
| | GLSZM | ZE | SAE | GLNU<br>SZNUN<br>ZP | GLNUN<br>GLV<br>HGLZE<br>LAE<br>LAHGLE<br>LALGLE<br>LGLZE<br>SZNU<br>SAHGLE<br>SALGLE<br>ZV |
| | GLDM | DE | | DNU<br>DNUN<br>SDE | DV<br>GLNU<br>GLV<br>HGLE<br>LDE<br>LDHGLE<br>LDLGLE<br>LGLE<br>SDHGLE<br>SDLGLE |

**Supplemental Table 10.** Effect of Attenuation Correction

| Feature category | Feature parent | COV≤5% | 5%<COV≤10% | 10%<COV≤20% | COV>20% |
|---|---|---|---|---|---|
| First order | FO | 10Percentile<br>Entropy | Kurtosis<br>Minimum | 90Percentile<br>Energy<br>IQR<br>Maximum<br>MAD<br>Mean<br>Median<br>Range<br>RMAD<br>RMS<br>Skewness<br>Uniformity | Total Energy<br>Variance |
| Second order | GLCM | CS<br>Idmn<br>Idn<br>Imc1<br>Imc2<br>Jent<br>SE | DEnt | Correlation<br>DAve<br>Id<br>JA<br>JEne<br>MP<br>SA | AC<br>CP<br>CT<br>Contrast<br>DVariance<br>Idm<br>IV<br>SS |
| Higher order | GLRLM | LRE<br>RE<br>RLNUN<br>RP<br>SRE | RLNU | GLNUN | GLNU<br>GLV<br>HGLRE<br>LRHGLE<br>LRLGLE<br>LGLRE<br>RV<br>SRHGLE<br>SRLGLE |
| | GLSZM | SAE<br>ZE | SZNUN<br>ZP | GLNU<br>GLNUN<br>SZNU | GLV<br>HGLZE<br>LAE<br>LAHGLE<br>LALGLE<br>LGLZE<br>SAHGLE<br>SALGLE<br>ZV |
| | GLDM | DE | DNU<br>SDE | DNUN<br>LDE | DV<br>GLNU<br>GLV<br>HGLE<br>LDHGLE<br>LDLGLE<br>LGLE<br>SDHGLE<br>SDLGLE |

**Discussion**

Radiomics is a new advanced approach for better disease management by using fast, non-invasive, easy, and cost effective methodology (10, 16). In this approach, features extracted from medical images are used for clinical applications and disease management (16, 22-26). However, it is important to note that radiomics suffers from fluctuation in the features against changing imaging settings, segmentation, and processing. Due to this, previous studies have suggested that radiomics features must be assessed in terms of repeatability, reproducibility, and robustness before applying them in clinical decision-making (27).

This study analyzed the reproducibility of cardiac SPECT radiomics features against changes in imaging settings including reconstruction, number of iterations, number of subsets, different filter, Gaussian-FWHM, cut–off, order, matrix size, attenuation correction, and number of views. Results showed that several features are reproducible while many of them are not. It was also found that the effects of different imaging settings are dependent on the type of setting and feature characteristics.

As shown in the heatmap, IDMN and IDN features from GLCM, RP from GLRLM, ZE from GLSZM, and DE from the GLDM feature sets were the only features that were most reproducible (COV≤5%) against change in all imaging settings. In addition, the IDMN feature from GLCM, LALGLE, SALGLE, and LGLZE features from GLSZM and the SDLGLE feature from GLDM feature sets were the only features that were less reproducible (COV>20 %) against changes in all imaging settings.

The results show that the matrix size has the greatest impact on feature variability, which is in concordance with previous studies. Previous study showed that the impact of matrix size changes in PET/CT radiomic features(21). After matrix size, the number of views has the large impact on radiomics future values.

Cardiac radiomics is a new approach for better CVDs management. Several studies have been conducted in attempts to address this issue. Ashrafinia *et al.* (28, 29), applied texture and radiomics analysis to clinical myocardial perfusion SPECT imaging to predict coronary artery calcification (CAC) from CT imaging. A study by Kolossváry *et al.*(30), showed that radiomics features are superior to conventional quantitative computed tomographic metrics in identifying coronary plaques with napkin-ring signs. Neisius *et al.*(31) examined the diagnostic ability of cardiovascular magnetic resonance image radiomics features in differentiating between hypertensive heart disease (HHD) and hypertrophic cardiomyopathy (HCM). Their study showed that native T1 imaging discriminates between HHD and HCM patients and provides incremental value over global native T1 mapping.

The results can be applied in various clinical settings before any decision making and to design and discover more imaging biomarkers. A wide range of studies have been done on the repeatability and reproducibility of radiomic features. Recently, Traverso *et al.*(32) analyzed which types of radiomic features have been shown to be repeatable/reproducible in peer-reviewed studies, and to what degree of repeatability and reproducibility might be achievable. However, this review does not mention the presence of research conducted on SPECT radiomics features repeatability and reproducibility. To the best of our knowledge, this current study is the first work on this topic and it's results are beneficial for researchers and clinicians working in this field (9, 12, 33).

Although these results are significant, this study has some limitations. This study was conducted by using phantom and more clinical studies are needed to explore the impact of biological factors on the radiomics features.

## Conclusion

This multi-scanner phantom study analyzed the reproducibility of Cardiac SPECT radiomics feature against changes in imaging settings including reconstruction, number of iterations, number of subsets, different filter, Gaussian-FWHM, cut–off, order, matrix size, attenuation correction, and number of views. Repeatability and reproducibility of SPECT/CT radiomics texture features in different imaging settings is feature-dependent. Additionally, different image acquisitions and reconstructions have different effects on radiomics texture features. Low COV radiomics features should be consider for further clinical studies.